# Exploring Potential Environmental Benefits of Asteroid Mining

Andreas M. Hein[a]*, Michael Saidani[a], Hortense Tollu[a]

[a] Laboratoire Genie Industriel, *CentraleSupélec, Université Paris-Saclay,* Gif-sur-Yvette, France
andreas-makoto.hein@centralesupelec.fr
* Corresponding Author

**Abstract**

Asteroid mining has been proposed as an approach to complement Earth-based supplies of rare earth metals and supplying resources in space, such as water. Existing research on asteroid mining has mainly looked into its economic viability, technological feasibility, cartography of asteroids, and legal aspects. More recently, potential environmental benefits for asteroid mining have been considered. However, no quantitative estimate of these benefits has been given. This paper attempts to determine if and under which conditions asteroid mining would have environmental benefits, compared to either Earth-based mining or launching equipment and resources into space. We focus on two cases: Water supply to cis-lunar orbit and platinum mining. First, we conduct a state-of-the-art of current environmental life cycle assessment for the space domain and platinum mining. Second, a first order environmental life cycle assessment is conducted, including goal and scope definition, inventory analysis, and impact assessment. We compare water supply to cis-lunar orbit with and without asteroid mining and go on to compare terrestrial with space-based platinum mining. The results indicate that asteroid water mining would have environmental benefits, as soon as the amount of water supplied via mining is larger than the mass of the spacecraft used for mining. For platinum mining, we find that by comparing the operations phase of terrestrial and space mining, space mining would have a lower environmental impact, if the spacecraft is able to return between 0.3 to 7% of its mass in platinum to Earth, assuming 100% primary platinum or 100% secondary platinum, respectively. For future work, we propose a more detailed analysis, based on a more precise inventory and a larger system boundary, including the production of the launcher and spacecraft.

## 1. Introduction

Mining asteroids, and in particular mining Near Earth Asteroids (NEAs) has been frequently proposed as a source of resources for space and terrestrial applications [1]–[3]. Two broad categories of resources can be distinguished: volatiles and metals. Ross [4] identifies a variety of applications for these resources such as construction, life support systems, and propellant. Volatiles such as water are of particular interest for in-space applications, due to their abundance in carbonaceous (C-type) asteroids and their relative ease of extraction. For example, Calla et al. [5] explore the technological and economic viability of supplying water from NEAs to cis-lunar orbit.

Regarding the supply of resources for terrestrial applications, only resources with a high market value are interesting, due to the high transportation cost. Hence, expensive metals such as rare earth metals and in particular the subgroup of platinum group metals have been the subject of asteroid mining studies [6]. The supply of platinum group metals is crucial for many terrestrial "green technologies" such as fuel cells and catalyzers [7]–[10]. However, there are two major concerns regarding platinum group metals. First, current supplies of platinum group metals are dominated by only a few countries, namely, South Africa, Russia, and Canada, which introduces political uncertainties into the supply chain [11]. The second concern is regarding the environmental impact of mining platinum group metals. Mines tend to go deeper and deeper, as resources in upper layers are depleted, which increases already high greenhouse gas emissions (currently ~40,000t $CO_2$ per ton of platinum) [11], [12]. Mitigating these issues has led to initiatives for recycling rare Earth metals and investigating substitutes [13]–[15]. In addition, the local environment is severely impacted due to the use of hazardous substances during the extraction process [11].

Despite the potential environmental benefits of asteroid mining, either by reducing the number of launches into space or moving terrestrial industries into space, no dedicated studies for exploring these benefits has been conducted to the authors' knowledge. Existing research on asteroid mining has mainly looked into its economic viability [2], [6], [16], [17], technological feasibility [2], [18]–[23], cartography of asteroids [24], [25], and legal aspects [26]–[28]. More recently Hennig [29] and MacWhorter [30] have introduced environmental arguments for asteroid mining, in particular with regards to platinum group metals. They refer to the benefits of asteroid mining for the environment and sustainability, but do not provide any analysis or quantitative backing.

This article addresses this research gap by providing an initial, first-order estimate of the potential environmental benefits of asteroid mining, exemplified via the case of water and platinum mining.

## 2. Literature survey

Two different research streams are relevant for an environmental life cycle assessment of asteroid mining: The environmental life cycle assessment of space systems and platinum.

*2.1 Space systems life cycle assessment*

The environmental life cycle assessment (LCA) of space systems is a rather recent domain. Environmental life cycle assessment is an approach for assessing the sustainability of products or systems. Chytka et al. [31] present an integrated approach to life cycle assessment, however, environmental aspects are not taken into account. Ko et al. [32] provide an overview of impacts of space activities on the space and Earth environment. They conclude that existing LCA approaches are insufficient for addressing impacts to space and suggest the development of additional impact categories. Neumann [33] applies LCA to launchers and provides a detailed inventory of inputs and outputs. However, the environmental impact from combustion exhausts is not taken into account. Austin et al. [34] present an overview of ESA activities on adopting LCA for space systems and mention their application to EcoSat, Ariane 5, Vega, Ariane 6, and four complete space missions. Wilson and Vasile [35] present a framework for integrating LCA into a concurrent engineering environment. De Santis et al. [36] present a methodology for a cradle-to-grave LCA for the European space sector and applied to the Astra 1N and MetOp A missions. The ESA Space system Life Cycle Assessment (LCA) guidelines [37] introduce an LCA approach based on the ISO 14040 / 14044 standard, tailored to the European space sector.

Although LCA has been applied to several case studies in the space domain, its introduction is recent and no application to asteroid mining could be found.

*2.2 Platinum mining life cycle assessment*

Platinum mining LCA studies are routinely performed by platinum mining companies, primarily for the estimation of their carbon footprint. The reported values are usually limited to greenhouse gas emissions and energy consumption. The global warming potential of emissions is commonly expressed in carbon dioxide equivalent or in short $CO_2eq$ over a period of 100 years [38], [39]. Although not a lot of detail is given for how the LCA is conducted, we assumed that a carbon footprint analysis of either Scope 1 (emissions are direct emissions from owned or controlled sources) or Scope 2 (indirect emissions from the generation of purchased energy) has been performed [40].

Several reports on the carbon footprint of primary platinum production exist, such as Bossi and Gediga [41], Montmasson-Clair [42], Cairncross [43], and by the Science Advice for the Benefit of Europe [44]. Glaister and Mudd [45] present qn extensive comparison of the environmental impact of platinum mining, based on $CO_2eq$ values reported by various platinum mining companies. $CO_2eq$ values for platinum from secondary production (recycled platinum) is available, for example, in the LCA database Impact 2002.

Saidani [12] estimates a mean value of 40 tons $CO_2eq$ of greenhouse gas emissions per kg of platinum from primary platinum production, based on a literature survey. For secondary production, a value of 2 tons $CO_2eq$ per kg of platinum is estimated. We will use these values as a reference.

## 3. Asteroid mining environmental life cycle assessment

We perform a first-order cradle-to-gate (extraction to factory gate) life cycle assessment of water and platinum asteroid mining, limited to greenhouse gas emissions.

*3.1 Goal and scope definition*

The scope and functional unit define the reference against which mining activities on Earth and space are compared.

The functional unit quantifies the service delivered by the product system. For our two cases of water mining in space and platinum mining on Earth and space, we use the following functional units:

- 1 kg of water delivered to cis-lunar orbit.
- 1 kg of platinum supplied to the Earth.

In terms of scope, we limit our analysis to greenhouse gas emissions, as data is available from various sources. Furthermore, our system boundary is drawn to include the operations phase, which includes E1, launch and commissioning phase, E2, utilisation phase in space, and F, disposal, according to the ESA lifecycle assessment guidelines [37]. Contrary to the guidelines, in our case we interpret F not as disposal but re-entry of platinum to Earth. For an Earth-based mine, the operations phase would essentially include the operation of the mine post installation. Furthermore, the boundary is drawn around the direct production and refining system of platinum or water. The reason for the limitation to the operations phase is that the publicly available sources of LCA data for platinum mining is limsted to the operations phase, which contains extraction and refining.

One could argue that for space-based mining, only E2 should be taken into consideration, as the production of the mining infrastructure is not taken into account for Earth-based mining. However, we interpret the launch infrastructure with launch pads, fuel depots, etc. as part of the infrastructure as well as launchers, and spacecraft. We therefore consider operations in the wider sense of operating this whole infrastructure, eoncompassing both E1 and E2. Consistent with carbon footprint analysis, we take Scope 1 (emissions are direct emissions from owned or controlled sources) and Scope 2 (indirect emissions from the generation of purchased energy) into account, in order to arrive at results that can be compared with platinum LCA results from the literature.

*3.2 Lifecycle inventory*

For the lifecycle inventory, fuel for the launcher and electricity for the launch infrastructure are considered as inputs. The output is limited to greenhouse gas emissions, for the simple reason that it is rather easy to find values for platinum mines. The values for electricity consumption for a launch of a Falcon Heavy-class rocket in Neumann [33] indicate that it is rather negligible compared to the greenhouse gas emissions from fuel combustion during ascent. Neumann [33] does not take greenhouse gas emissions from fuel combustion into account. However, we use the LCA conducted for kerosene by [46], where the greenhouse gas emissions from combustion is the dominant contribution to greenhouse gas emissions in the kerosene supply chain. In the following, we use a rough value of 3 kg $CO_2$eq per kg of Kerosene combusted.

*3.3 Bootstrapping factor*

We use the bootstrapping factor *b* as a figure of merit, which we define as kg of payload mass launched into space vs. kg of resources delivered to the target destination.

$$b = \frac{m_{res}}{m_{pl}} \quad (1)$$

$m_{res}$ indicates the mass of resources mined and supplied to the target destination and $m_{pl}$ the mass of the payload launched into space for the mining operation.

For the case of water, the bootstrapping factor allows for a comparison between launching water from Earth and supplying mined water to a target destination. For example, a 500 kg spacecraft (wet mass) is launched into space for mining an asteroid and the spacecraft delivers 1000 kg to its target destination, the bootstrapping factor is 2. When a 500 kg spacecraft carrying water is launched into space, delivering 200 kg of water to its target destination. *b* is 0.4. Comparing the water asteroid mining example with direct water delivery yields a ratio of 5. In order to make environmental sense, *b* for mining has to be larger than the *b* for direct water

delivery. In the example above, this means $b > 0.4$. We can therefore write:

$$\frac{b_{mining}}{b_{direct}} = \frac{m_{res\_mining}}{m_{res\_direct}} > 1 \quad (2)$$

For linking *b* with environmental impact on Earth, a multiplier needs to be added, which converts the payload mass in a destination in space with a common payload reference, such as payload to LEO. Using the ratio from (2) and introducing the mass-specific environmental impact $\varepsilon$ yields the following equation for the mass-specific environmental impact of asteroid mining, compared to the direct delivery of a resource.

$$\varepsilon_{mining} = \frac{b_{direct}}{b_{mining}} * \varepsilon_{direct} = \frac{\varepsilon_{launch}}{b_{mining}} \quad (3)$$

$\varepsilon_{direct}$ indicates the mass-specific environmental impact of direct delivery of a resource. $\varepsilon_{launch}$ is the mass-specific environmental impact during launch. Equation (3) is not only valid for water mining but also for the case of mining and returning resources from space to Earth, such as platinum. For the latter, $\varepsilon_{mining}$ needs to be smaller than $\varepsilon_{mining\_Earth}$, the mass-specific environmental impact of mining on Earth:

$$\varepsilon_{mining} < \varepsilon_{mining\_Earth} \quad (4)$$

**3. Results**

*3.4 Asteroid water mining*

For the nominal case of supplying water to cis-lunar orbit, the environmental impact of launching 1 kg of water from Earth to a cis-lunar orbit is calculated.

Based on a previous analysis for asteroid water mining in Calla et al. [5] and Hein & Matheson [16], a range for $b_{mining}$ can be estimated between 0 and lower two-digit numbers. We calculate a lower bound for the $CO_2$eq of launching 1 kg of water to cis-lunar orbit, which only includes the $CO_2$ released from the combustion of kerosene, using Falcon Heavy data from Spaceflight 101 [47]. About 30 kg of kerosene is burned per kg of payload to cis-lunar orbit. We multiply this value by 3 kg $CO_2$eq per kg of kerosene burned, a factor introduced in 3.2. We therefore get a value of 90 kg of $CO_2$eq per kg of water delivered to cis-lunar orbit as a lower bound. Furthermore, it is assumed that all of the kerosene of the first stage and boosters are burned within the Earth's atmosphere. It is assumed that the second stage has no impact in terms of $CO_2$ emissions on Earth's atmosphere.

Using the bootstrapping factor $b_{mining}$, we get $CO_2$eq values for the case where an asteroid mining spacecraft is launched and returns per kg of spacecraft mass b-times its mass. Table 1 shows the resulting values. It can be seen that substantial savings in greenhouse gas emissions can be achieved.

Table 1: $CO_2$eq values for delivering 1 kg of water to cis-lunar orbit, with respect to the bootstrapping factor b

| $b_{mining}$ | $CO_2$eq [kg per kg of water] only Kerosene |
|---|---|
| 1 | 90 |
| 5 | 18 |
| 10 | 9 |
| 20 | 4.5 |
| 30 | 3 |
| 40 | 2.3 |

*3.5 Asteroid platinum mining*

For the case of platinum, Earth-mining is the reference and the impact of returning platinum to Earth needs to be taken into consideration. It is known that during re-entry a spacecraft releases $H_2O$ and NOx in the Earth's upper atmosphere via the re-entry shock wave and material released via ablation [48], [49]. $N_2O$ has a global warming potential of between about 265–298, 310 times $CO_2$ [39], [50].

Park and Rakich [51] estimate that about 17.5±5.3% of the Space Shuttle mass is released in the form of NOx during re-entry. As a conservative estimate, we use 20% and assume that predominantly $N_2O$ is released. Furthermore, we assume that for 1 kg of platinum, about 1 kg of additional mass is required for re-entry (heatshield, GNS, parachute etc.). Hence, roughly 0.2 kg of $N_2O$ is released per kg of platinum returned to Earth, which translates into roughly an equivalent of 60 kg of CO.

As a result, we get a total kg $CO_2$eq per kg Pt of 150 kg. Given various uncertainties, we see that the total $CO_2$eq of an asteroid mining mission is on the order of dozens to hundreds of kg $CO_2$eq per kg of platinum returned.

If we compare these rough estimates with the $CO_2$eq values for Earth-based platinum mining, we immediately see that the global warming effect of Earth-based mining is several orders of magnitude larger, even for secondary platinum. Table 2 shows the ratio between the Earth-based platinum mining emissions and the space-based mining emissions. A difference of two orders of magnitudes for primary platinum and one order of magnitude for secondary platinum is observed. For a mixture of primary and secondary platinum, we get values with two orders of magnitude difference.

Table 2: Comparison of space and Earth-based platinum mining greenhouse gas emissions

| $b_{mining}$ | $CO_2$eq / kg Pt | Ratio Earth reference (40 t / kg $CO_2$eq) vs. space | Ratio Earth reference (2 t / kg $CO_2$eq) vs. space | Earth: 33% secondary, 66% primary platinum vs. space |
|---|---|---|---|---|
| 1 | 150 | 267 | 13 | 182 |
| 5 | 78 | 513 | 26 | 350 |
| 10 | 69 | 580 | 29 | 396 |
| 20 | 65 | 620 | 31 | 424 |
| 30 | 63 | 635 | 32 | 434 |
| 40 | 62 | 643 | 32 | 439 |

Although the $CO_2$eq values used for space-based platinum mining represent a lower bound, we can estimate that even one order of magnitude higher emissions would lead to one order of magnitude savings, compared to Earth-based mining.

*3.6 Carbon tax effects*

A straight-forward consequence of greenhouse gas emissions is that they have an economic effect, once carbon tax is introduced. Given the large differences in $CO_2$eq emissions for Earth and space-based platinum mining, Earth-based mining would be penalized with the introduction of carbon tax. As shown in Table 3, using the value of 50 t of $CO_2$eq per kg of platinum mined in 2030, extrapolated from its current value of 40 and a conservative carbon tax value of €70 per ton, we obtain a carbon tax of €3,500 per kg of platinum. Given today's price levels for platinum and assuming that these remain similar, a penalty of ~10% needs to be added on top of the cost of Platinum production. Currently, the platinum mining industry is operating at low profit margins or even at a loss. The 10% tax could be compensated via a higher efficiency of the mining process and a potentially higher degree of renewable energy sources for electricity supply, as the majority of greenhouse gas emissions are generated by burning hard coal, at least in the case of South Africa [41]. However, it is unclear how much platinum mining companies might influence decisions that concern the energy mix on a country level.

Table 3: Estimates of carbon tax for platinum production

| Year | t CO$_2$eq/kg Pt | Carbon Price (€) | Carbon Tax (€) / kg Pt |
|---|---|---|---|
| 2017 | 40 | 5 | 200 |
| 2030 | 50 | 70 | 3500 |
| 2050 | 60 | 120 | 7200 |

## 5. Discussion

The results of the asteroid mining LCA show that for a broad range of bootstrapping factors $b_{mining}$, substantial environmental benefits could be reaped for both, water and platinum mining. The range of $b_{mining}$ is consistent with the values for $b_{mining}$ presented in Hein and Matheson [16] and should cover realistic mining scenarios. As with LCA in general, this result depends on the initial scope of the assessment.

There are several limitations to the analysis presented in this article. For example, the environmental impact of rocket launchers could be reduced by applying eco-design principles, such as the use of "green propellants", the reuse of components such as rocket stages etc. Some of these options are discussed in Neumann [33].

Furthermore, only greenhouse gas emissions have been considered, and a more extensive LCA would require the consideration of further impact categories. Of particular relevance for launchers is ozone layer depletion, as combustion products are directly released above the troposphere. Hence, adding midpoint and endpoint impact categories would create a more complete picture of the environmental impact of an asteroid mining mission. However, at least for the case of platinum mining, we are limited by the availability of LCA data beyond CO$_2$eq and energy consumption.

Another limitation is that emissions from spacecraft operations have not been considered. Sending 1 kg of water into cis-lunar orbit takes less time than an asteroid mining mission, a few days versus hundreds of days to years. Emissions from ground station operations are proportional to the duration of the mission and could change the result in favour of launching water.

The impact of off-nominal behaviour has also not been considered. For example, the environmental impact of a failed launch within the Earth atmosphere would be much larger than for a successful launch, as the entire propellant would be burned within the atmosphere, including that of the upper stage(s).

A topic that merits further investigation is the assessment of the in-space impact of asteroid mining. Such an assessment could be extended to trade-offs between terrestrial and space impact. The recent literature on in-space impact assessment could provide a starting point [32], [52], [53].

## 6. Conclusions

This article provides a first-order analysis of the potential environmental implications of asteroid mining, with a focus on greenhouse gas emissions. We introduce a bootstrapping factor, the ratio of resources delivered to the target destination and the payload mass launched into space that allows for a comparison of various asteroid mining missions. The results for the case of in-space water supply and platinum mining indicate that for typical values of the bootstrapping factor, asteroid mining generates substantial environmental benefits compared to its alternatives.

For future work, a more extended LCA for asteroid mining missions would provide a more extensive picture of its environmental impacts. Further, combining economic and environmental assessment seems to be promising for identifying mining architectures that show a good performance with respect to both criteria. Another interesting topic would be a framework for conducting trade-offs between terrestrial and in-space environmental impacts such as the generation of space debris and the occupation of orbits.